\documentclass{ws-p8-50x6-00}

\begin{document}

\title{NUCLEAR CHRONOMETERS}

\author{J. J. Cowan}

\address{Department of Physics \& Astronomy, University of Oklahoma,
Norman, OK \\  73019, USA\\E-mail: cowan@mail.nhn.ou.edu}

\author{C. Sneden}

\address{Department of Astronomy, University of Texas,
Austin, TX
78712, USA\\E-mail: chris@verdi.as.utexas.edu}

\author{J. W. Truran}

\address{Department of Astronomy, University of Chicago,
Chicago, IL
60637, USA\\E-mail: truran@nova.uchicago.edu}

\maketitle

\abstracts{
Observations of metal-poor Galactic halo stars indicate 
that the abundance pattern of the (heaviest) neutron-capture elements
is consistent with the scaled solar system {\it r}-process abundances. 
Utilizing  the radioactive ({\it r}-process) element thorium,  age determinations
have been made for several of these same stars,  
placing constraints
on both Galactic and cosmological age estimates.}

\section{Introduction}
Stellar observations of neutron-capture elements ({\it i.e.}, those produced
in the slow- or rapid-neutron capture processes) provide a number of 
important clues about   the formation history, chemical evolution and age of 
these elements in the Galaxy.  
In particular, the abundances of the {\it n}-capture elements in metal-poor 
({\it i.e.}, low iron abundance)
Galactic halo stars can be employed  to determine the nature of the 
progenitors
and the nucleosynthesis history in  the early Galaxy.
Further, the 
changes in the stellar abundance trends with respect to  metallicities
(and times) can provide clues  about the nature of Galactic chemical
evolution and the sites for neutron-capture synthesis. 
Just as importantly, certain    
long-lived radioactive 
elements, 
such as thorium and uranium, are  produced entirely in the {\it r}-process. 
The abundance levels of these  nuclear chronometers  
in the most metal-poor halo stars provide  direct  age 
determinations 
and hence
set lower limits on Galactic and cosmological age estimates.  

\section{Stellar $n$-Capture Abundance Distributions}
\par
There have been a number of recent ground-based (see {\it e.g.} 
Sneden {\it et al.}\cite{sne96}, 
McWilliam\cite{mcw98}, Burris {\it et al.}\cite{bur00}, 
Sneden {\it et al.}\cite{sne00}) 
and space-based (Cowan {\it et al.}\cite{cow96}, 
Sneden {\it et al.}\cite{sne98}) 
observations of metal-poor Galactic halo stars. 
Several of these studies have concentrated on the star CS 22892--052 
(Cowan {\it et al.}\cite{cow95}, Sneden {\it et al.}\cite{sne96}, 
Sneden {\it et al.}\cite{sne00}).
This star,  while ultra-metal-poor ([Fe/H] = --3.1), 
is neutron-capture rich with an average
abundance level of 30-40 times solar (with respect to iron) for the 
elements heavier than Ba. These large overabundances have allowed the 
detection of a number of elements for the very first time in 
very metal-poor halo  stars. 
In Figure 1 (at the top) the neutron-capture abundances
in CS 22892--052 for elements with atomic number,  Z, $\ge$ 56  are shown.
Also shown in comparison is the solar system elemental 
{\it r}-process abundance 
distribution, indicated by a solid line. 
This distribution has been determined by 
summing the individual isotopic contributions from the s- and the 
{\it r}-process in solar system material, as determined by 
K{\"a}ppeler {\it et al.}\cite{kap89}  from neutron-capture cross section
measurements. 
(See Burris {\it et al.}\cite{bur00}  for  more details.) 
It is clear in Figure 1 that there is a remarkable agreement between
the elemental abundances in this star and the scaled solar system 
{\it r}-process 
abundances -- the relative abundances of the heavy {\it n}-capture elements
are similar in CS 22892--052 and the solar system. 
While this argument has been made previously,
the detection of additional elements in CS 22892--052 
has strengthened that argument.

While the agreement between scaled solar system {\it r}-process abundances and
CS 22892--052 is very suggestive, it might 
be considered coincidental if it was a unique situation. This is not 
in fact the case. Recent observations of other neutron-capture rich 
halo stars have shown a similar pattern. The {\it n}-capture abundances in 
one such star, HD 115444,  with a  metallicity  
similar to CS 22892--052,  
is shown in the lower part of Figure 1 (Westin {\it et al.}\cite{wes00}). 
The abundances of the elements 
in this star have been arbitrarily (uniformly) 
reduced (i.e., moved downward) for 
illustration purposes - the actual abundances are approximately a factor of 
5 lower than in CS 22892--052. 
The agreement between the abundances 
in HD 115444 and  the scaled solar system elemental
{\it r}-process abundance curve
is again excellent and extends through the heaviest
stable elements including Os and Pt in the so-called 3$^{\rm rd}$ 
{\it r}-process peak. 
Still further support for this agreement  
is provided by recent observations of 
stars in the globular cluster M15. The abundances of the
heaviest neutron-capture
elements in three giants in that cluster are again consistent with a scaled
solar system {\it r}-process abundance distribution 
(Sneden {\it et al.}\cite{sne0b}). 
\begin{figure}[t]
\epsfxsize=20pc 
\hspace*{0.5in}
\epsfbox{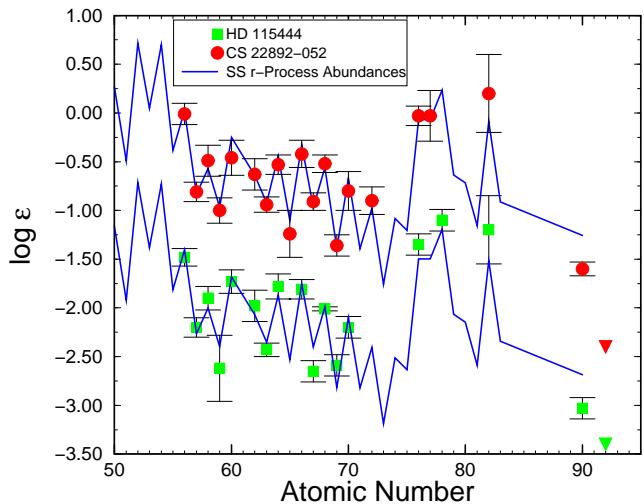} 
\caption{The heavy {\it n}-capture abundance patterns for the two 
stars CS 22892--052  and 
HD~115444 are compared with the solar system {\it r}-process 
abundance distribution
(solid line) . 
\label{figure1}}
\end{figure}

It is also apparent from Figure 1 that elements, such as barium,
in CS~22892--052 and HD~115444 that are typically 
thought of as {\it s}-process elements (and predominantly produced by that 
process in solar system material -- see Burris {\it et al.}\cite{bur00}) were 
almost exclusively produced by the {\it r}-process early in the history of
the Galaxy. This result affirms earlier predictions (Truran\cite{tru81})
concerning  the dominance of  {\it r}-process nucleosynthesis early in
the history of the Galaxy. 
The detection  of {\it r}-process material in the metal-poor and old halo stars,
including the heaviest {\it r}-process  elements, 
indicates an early presence of these elements in the Galaxy. That in
turn  also places constraints on the type of progenitors of the {\it r}-process
elements and, in particular, suggests  
rapidly evolving progenitors of the halo stars. 
This follows since the early Galaxy appears to be chemical inhomogeneous 
in n-capture elements 
(see Sneden {\it et al.} in this volume) implying a relatively  
short (with respect to stellar evolutionary) timescale between the 
death of the progenitors and 
the formation of the halo stars.

Until now the element domain between
Zr and 
the heavier {\it n}-capture elements,
{\it i.e.}  40 $<$ Z $<$ 56,  
has been largely unexplored.
For the first time Sneden {\it et al.}\cite{sne00}  have detected six elements  
in this region in CS 22892--052, as shown
in Figure 2. 
It is clear from the abundance comparisons in 
Figure 2, however,  that the lighter
neutron-capture elements, specifically the six newly detected elements 
in CS 22892--052,  
do not fall on the same curve as the heavier neutron-capture elements.
These recent observations lend support to earlier 
suggestions - based upon analyses of solar system meteoritic material - 
of a second {\it r}-process with the heavier elements (Ba and above)
produced in more rapidly occurring events and the lighter elements in 
less commonly occurring syntheses (Wasserburg {\it et al.}\cite{was96},
Wasserburg and  Qian\cite{was00}; see  
Sneden {\it et al.}  in this volume for discussion.) 

\begin{figure}[t]
\epsfxsize=20pc 
\hspace*{0.5in}
\epsfbox{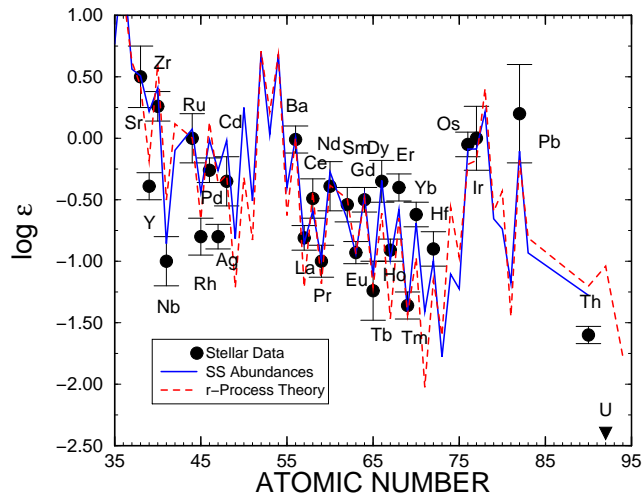} 
\caption{The total neutron-capture abundance pattern 
in CS 22892--052 compared with 
the solar system {\it r}-process abundances (solid line) and a theoretical 
{\it r}-process abundance curve (dashed line). 
\label{figure2}}
\end{figure}
\section{Ages from Cosmochronometry}
The long-lived radioactive nuclei (known as chronometers)
in the thorium-uranium region
are formed entirely in the {\it r}-process and can be used to determine the
ages of
stars and the Galaxy. 
The technique of using the observed stellar thorium, particularly 
as a ratio to a stable element, to obtain ages  
was first introduced by Butcher\cite{but87}
and developed by Fran\c cois {\it et al.}\cite{fra93}.
The detection of thorium in CS 22892--052 and 
HD 115444 (see Figure 1), therefore, allows for chronometric stellar age 
estimates.  
The age is derived by comparing the observed stellar abundance ratio
Th/Eu  
with predictions of the  initial (zero-decay) value of
Th/Eu at the time of the formation of these elements.
(Eu is employed since it is also a predominantly an {\it r}-process element and 
has spectral features that are easily observed
in most metal-poor stars.) 
Cowan {\it et al.}\cite{cow99} and Westin {\it et al.}\cite{wes00} 
compared the  observed values of Th/Eu in  CS~22892-052 and
HD~115444 
with that in solar system material to
obtain an average age of 13.8 Gyr. This estimate is in fact a lower limit 
on the age of these very low metallicity stars. This follows  
since 
the solar Th/Eu ratio (at the formation of the
solar nebular 4.5 Gyr  ago) represents a lower limit to the
zero decay-age {\it r}-process abundances -- Eu is stable and Th (although
constantly produced and ejected into the interstellar medium) is partially
decayed.

Further refinements in the age  estimates 
can be obtained by determining  the zero-decay
abundance of thorium when produced in an {\it r}-process site. 
This,  however, depends critically upon theoretical 
predictions for  very neutron-rich nuclei, whose  properties are 
in general not obtainable by experimental determination.   
We can, however,  employ the stable abundances of the 
neutron-capture elements  and the solar system {\it r}-process abundances  
to constrain the theoretical calculations and,  hence,  assess the 
reliability of any  
nuclear mass formulae utilized in those calculations.  
One such calculation, based upon the work of 
Pfeifer {\it et al.}\cite{pfe97}
(see also Cowan {\it et al.}\cite{cow99} for details),  
is shown in Figure 2 by a dashed line.
It is seen in the figure that this calculation gives  very good
agreement for the abundance predictions of the heavy {\it n}-capture elements 
in CS 22892--052 and the solar system  abundances. In this same 
calculation, constrained by the stable abundances, abundance 
predictions for the radioactive heavy elements are simultaneously
derived. 
There is still an uncertainty associated with the abundance 
predictions for the radioactive chronometers thorium and uranium.
Unlike the stable heavy solar system elements, such as Pt,
which can be compared directly with theoretical predictions, no such
comparison is possible with Th or U.
However, 
the $\alpha$-decay of heavier nuclei (209 $<$ A $<$ 255), 
including those from Th and U,   
is responsible for the production of the stable lead
and bismuth isotopes, whose abundances are known and thus can be 
compared with
theoretical estimates.
Employing this important solar system abundance constraint on the theoretical
{\it r}-process calculations to determine the zero-decay Th/Eu abundances,
Cowan {\it et al.}\cite{cow99} 
and Westin {\it et al.}\cite{wes00} found an average age for CS 22892--052 and
HD 115444 of 15.6 Gyr, with an estimated uncertainty of $\simeq$ 4 Gyr. 
Sneden {\it et al.}\cite{sne0b} report on thorium detections
in several stars in the globular cluster M15 and  obtain ages, $\simeq$ 14 
Gyr,  that
are also consistent with those found for the two halo stars.

While there are still uncertainties associated with the nuclear physics 
predictions,   
this technique offers great promise for stellar dating.
One large advantage is that  
such radioactive age determinations of   very low metallicity stars
are independent of, and thus avoid the inherently large uncertainties in,
Galactic chemical evolution models.
With the  addition of  more observations and better nuclear physics predictions,
the current radioactive age uncertainties should   be reduced.  
Finally,  we  note that while uranium has not been observed so far 
in either CS 22892--052 or
HD 115444 (upper limits on U are indicated by inverted triangles in Figures 
1 \& 2),
in the near future it may be possible to observe this element in some 
metal-poor halo stars. The addition of this other nuclear chronometer 
would help to constrain further
age estimates for the Galaxy and the Universe.

\section*{Acknowledgments}
We thank all of our colleagues who have collaborated with us on various
studies of $n$-capture elements in halo stars.
This research has received support from NSF grants 
AST-9986974 to J.J.C., 
AST-9987162 to C.S. 
and
from DOE contract B341495 to J.W.T., and from the Space
Telescope Science Institute grant GO-8342.


\begin{thebibliography}{99}

\bibitem{sne96} C. Sneden {\it et al.}, 
 {\it ApJ} {\bf 467}, 819 (1996).

\bibitem{mcw98}
A. McWilliam,   {\it AJ} {\bf 115}, 1640 (1998).

\bibitem{bur00}
D. L. Burris {\it et al.}, {\it ApJ} {\bf 544}, 302 (2000).

\bibitem{sne00} C. Sneden {\it et al.}, 
{\it ApJ} {\bf 533}, L139 (2000).

\bibitem{cow96} J. J. Cowan  {\it et al.}, 
{\it ApJ} {\bf 460}, L115 (1996).

\bibitem{sne98} C. Sneden {\it et al.}, 
{\it ApJ} {\bf 496}, 235 (1998).

\bibitem{cow95}
J. J. Cowan {\it et al.}, 
{\it ApJ} {\bf 439}, L51 (1995).

\bibitem{kap89}F. K{\"a}ppeler {\it et al.}, {\it Rep. Prog. Phys.} 
              {\bf 52}, 945 (1989).

\bibitem{wes00}J. Westin {\it et al.}, {\it ApJ} {\bf 530}, 783 (2000).

\bibitem{sne0b} C. Sneden {\it et al.}, 
 {\it ApJ} {\bf 536}, L85    (2000). 

\bibitem{tru81} J. W. Truran, {\it A\&A} {\bf 97}, 391 (1981).

\bibitem{was96} G. J. Wasserburg {\it et al.},  {\it ApJ} {\bf 466},
L109 (1996).

\bibitem{was00}
G. J. Wasserburg and Y.-Z. Qian, {\it ApJ} {\bf 529}, L21 (2000).

\bibitem{but87}
H. R. Butcher, {\it Nature} {\bf 328},  127 (1987).

\bibitem{fra93}
P. Fran\c cois {\it et al.},  {\it A\&A} {\bf 274}, 821 (1993).

\bibitem{cow99}
J. J. Cowan {\it et al.}, {\it ApJ} {\bf 521}, 194 (1999).

\bibitem{pfe97}
B. Pfeiffer {\it et al.},  {\it Z. Phys. A} {\bf 357}, 235 (1997).


\end{thebibliography}
\end{document}